\documentclass[a4paper]{article}
\usepackage{Odyssey2018}
\usepackage{epsfig,amssymb,amsmath}
\usepackage{multirow}
\usepackage{makecell}

\newcommand{\vect}[1]{\boldsymbol{\mathrm{#1}}}
\newcommand{\matr}[1]{\boldsymbol{\mathrm{#1}}}
\newcommand{\norm}[1]{\left\lVert#1\right\rVert}

\ninept

\title{On deep speaker embeddings for text-independent speaker recognition}

\makeatletter
\def\name#1{\gdef\@name{#1\\}}
\makeatother

\name{{\em Sergey Novoselov$^{1,2}$, Andrey Shulipa$^1$, Ivan Kremnev$^2$, Alexandr Kozlov$^2$, Vadim Shchemelinin$^1$}}

\address{
  $^1$ITMO University, St.Petersburg, Russia \\
  $^2$STC-innovations Ltd., St.Petersburg, Russia\\
\\
  \small \tt {\{novoselov,shulipa,kremnev,kozlov-a,shchemelinin\}@speechpro.com}}

\begin{document}
\maketitle

\begin{abstract}
We investigate deep neural network performance
in the text-independent speaker recognition task.
 We demonstrate that using angular softmax activation 
 at the last classification layer of a classification neural network instead of 
 a simple softmax activation allows to train a more generalized discriminative 
 speaker embedding extractor. 
 Cosine similarity is an effective metric for speaker verification in this embedding space.
 We also address the problem of choosing an architecture for the extractor.
 We found that deep networks with residual frame level connections  outperform  wide 
 but  relatively  shallow  architectures.
 This paper also proposes several improvements for
 previous DNN-based extractor systems to increase the speaker recognition accuracy.
We show that the discriminatively trained similarity metric learning approach 
outperforms the standard LDA-PLDA method as an embedding backend.
 The results obtained on Speakers in the Wild and NIST SRE 2016 evaluation sets demonstrate robustness of the proposed systems when dealing with close to real-life conditions.

\end{abstract}

\section{Introduction} \label{sec:intro}

Text-independent speaker recognition continues to be a challenging task for modern voice biometrics systems. Complex speaker voice information must be captured from highly variate data with no evident speaker patterns. Candidate solutions must generalize well in order to be applicable to new possible deployment conditions.
This work investigates prominent techniques from speaker recognition field combined with face recognition and general deep learning science to bring new thoughts on how speaker recognition systems can be developed.

I-vector-based systems are well known to be state-of-the-art solutions to the text-independent speaker verification problem \cite{kenny-fa, novel, novoselov_non_linear_plda}.
The i-vector framework has inspired deep learning system design in this field. Particularly, in studies \cite{novel, novoselov_dnn} they use an ASR deep neural network (ASR DNN) to divide acoustic space into senone classes, and the classic total variability (TV) model is applied to discriminate between speakers in that space \cite{kenny-fa}.
In such phonetic discriminative DNN-based systems two major techniques can be distinguished. Firstly, one can use DNN posteriors to calculate Baum-Welch statistics, and secondly, bottleneck features extracted with a network can be used in pair with speaker specific features (MFCC) for a full TV-UBM system training.

Deep learning is a powerful tool for data analysis with complex data distribution \cite{krizhevsky2012imagenet,szegedy2015going,kingma2013auto,asoftmax},
and many researches consider training deep non-linear extractors as a solution to the direct speaker discrimination task. There are several solid studies on advantageous usage of deep end-to-end solutions for discriminating speakers directly in a text-dependent task \cite{microsoft, google, bhattacharya2016deep}. 
Paper \cite{dnnsmall} describes a deep network that extracts small speaker footprints which are used to discriminate between speakers.

Paper \cite{snyder2017deep} presents a rather powerful implementation of a DNN speaker embedding extractor based on the speaker discriminative approach in the text-independent setting. One of the key features of the system proposed by its authors is the time-delay neural network architecture of the extractor \cite{peddinti2015time} with a statistics pooling layer designed to accumulate speaker information from the whole speech segment into a single vector, which they call an x-vector. X-vectors expose similarities to i-vectors from a total variability space, and so can be effectively used with the standard Linear Discriminant Analisys (LDA) followed by Probabilistic Linear Discriminant Analisys (PLDA) backend for solving the speaker recognition task. The authors show that a discriminatively trained x-vector extractor surpasses the standard i-vector extractor in terms of speaker recognition error on the challenging NIST 2016 test.
    
This study also speculates on the discriminative approach for learning a deep speaker embedding extractor in the text-independent scenario.
This problem is usually solved by training a deep neural network speaker classifier on a closed speaker set and using its bottleneck features as speaker embeddings after. These embeddings are believed to contain informative speaker factors and respond moderately to any other speech variation if trained effectively.

A feature extractor trained this way must generalize well in order to be applicable to an open-set speaker verification.
It is challenging to achieve this property only using standard first order optimization algorithms and standard cross-entropy loss for softmax outputs of the network.
The main reason for that is quick overfitting of such a system to corpus-specific speech features like the language, the audio channel type and quality and so on. Unfortunately, common regularization techniques like weight decay and dropout do not help much to resolve these issues.

However, according to our observations, more complex optimizators can be a solution to the problem, e.g. natural gradient, as was done for the x-vector extraction procedure. They allow the algorithm to converge to a lower optimum while being trained for multiple epochs.

Another option for building a more generalized extractor is a better choice of the loss function objective for the training stage and appropriate design of the architecture of a neural network.
  
Thus in this study we ponder on the right architecture for the neural network and a proper choice of the loss function. We take some ideas from the field of face recognition and general deep learning to our advantage and train a speaker embedding extractor that is compatible with the x-vector system but is computationally lighter due to the lower number of parameters.

In addition, we study performance of the x-vector-based system with a learnable Cosine Similarity Metric (CSML) backend \cite{Nguen}, which outperforms the standard LDA-PLDA backend for x-vectors.

The rest of the paper is organized as follows. Section \ref{sec:related} describes previous systems and techniques that we use as baseline or as a component of our systems. Section \ref{sec:deep_systems} describes highlights of each deep extractor we train and study along with their configuration presented in Tables \ref{tab:maxdnn}, \ref{tab:resnet}. Section \ref{sec:frontend_backend} describes details of the whole models, including frontend feature configuration and backend embedding classifiers. Sections \ref{sec:setup} and \ref{sec:results} cover experimental setups and results, and Section \ref{sec:discuss} speculates on strong points of each system and directions for further development of text-independent speaker recognition systems.

\section{Related work}
\label{sec:related}

This section describes state-of-the-art speaker recognition systems, including i-vector baselines and a DNN-based speaker embedding extractor.

\subsection{Baseline i-vectors}
\label{sec:dnn_posterior}

Most of the text-independent speaker recognition systems are based on the i-vector extraction framework. Typically, i-vector computation process can be decomposed into three stages: collection of sufficient statistics, extraction of i-vectors and a probabilistic linear discriminant analysis (PLDA) backend \cite{novel,kenny-fa,novoselov_dnn}.
Sufficient statistics are collected by using a sequence of feature vectors, e.g. melfrequency cepstral coefficients (MFCC), and are usually represented by Baum-Welch statistics obtained with respect to a GMM, which is called a universal background model (UBM). These statistics are converted into a single low-dimensional feature vector --- an i-vector --- that contains substantial information about the speaker and all other types of speech variability. After i-vectors are extracted, a PLDA model is used to produce verification scores by comparing i-vectors extracted from different speech segments. 

An alternative i-vector framework is based on deep neural networks. DNN/i-vector frameworks provides the best speaker recognition performance in "clean" speech conditions \cite{novel,sadjadi2016ibm}.
In the DNN-based i-vector framework a deep neural network substitutes a UBM in calculation of Baum-Welch statistics, followed by total variability factor analysis. Alternatively, DNN can be used for extracting bottleneck (BN) features, which are used together with MFCC in the standard UBM/i-vector framework. This also gives impressive speaker recognition performance \cite{sri_bn}.

A drawback of the latter systems is their low robustness when applied in real acoustic settings. This is confirmed with the NIST 2016 SRE results \cite{nist2016eval}.
In such setting standard acoustic i-vector is preferred.

In our experiments we use only DNN posterior-based i-vector extraction procedure for baseline.

\subsection{DNN speaker embeddings}

Impressive results can be obtained by using solely a deep neural network classifier as an extractor of speaker embeddings. Trained on basic speech features like log mel power signal spectrum to discriminate between speaker classes, such network is a convenient tool for extracting high-level speaker features from top levels of the network.

One example of such an extractor is the x-vector system \cite{snyder2017deep}. The systems uses MFCC features of the speech signal of a fixed as input features for a TDNN-type neural network. Each time-delay layer of the network gradually develops speaker-informative features on the feature map while also expanding context of a frame which is crucial for global speaker capturing.
Time-delay layers are followed by a stats pooling layer, which folds features along the time axis by calculating mean and standard deviation statistics of those features. As a result, we get a highly representative feature vector containing global speaker information. To get speaker embedding right, this vector is passed through several fully-connected layers of a classifier network. 

It is shown that x-vector-based systems outperform i-vector-based systems when applied to in-the-wild conditions \cite{snyder2017deep, snyderx}.
However, using softmax cross entropy loss function for extractor training does not allow to use standard metrics, such as cosine metric, for embedding scoring. In this case, an LDA-PLDA backend gives good results.

The detailed configuration for the x-vector extractor is presented in \cite{snyder2017deep}.

\subsection{Angular softmax loss with margin}


There are known many heuristics for effective discriminative learning of a deep extractor from the face recognition field. Most widely used regularized loss functions are centerloss with softmax loss \cite{centerloss}, contrastive loss \cite{chopra2005learning} and triplet loss \cite{hoffer2015deep,schroff2015facenet}. 
We believe that a robust speaker recognition system can be designed to extract embeddings onto a hypersphere manifold with a cosine similarity metric suitable for their comparison. An example of this representation is the commonly used i-vector normalization technique.
Work \cite{asoftmax} demonstrates an elegant way to obtain well regularized loss function by forcing learned features to be discriminative on a hypersphere manifold. This idea directly leads to a rigorous formulation of the angular margin softmax loss:

\begin{align*}\label{eq:asoftmaxloss}
L_\text{ang}=  \dfrac{1}{N}&\sum_{i}-\log(\\
&\dfrac{e^{\|x_i\|\cos{(m\theta_{i,y_i})}}}{e^{\|x_i\|\cos{(m\theta_{i,y_i})}} + \sum_{i\neq y_i}e^{\|x_i\|\cos{(m\theta_{i,y_i})}}})\\
\end{align*}
where  $N$ is the number of training samples $\{x_i\}_{i=1}^N$ and their labels $\{y_i\}_{i=1}^N$, $\theta_{i,y_i}$ is the angle between $x_i$ and the corresponding column $y_i$ of the fully connected classification layer weights $\matr{W}$, and $m$ is an integer that controls the size of an angular margin between classes. 

We use Asoftmax as an effective discriminative objective for training our deep embedding extractor.

\section{Analisys of deep learning systems}
\label{sec:deep_systems}

Architecture choice is crucial for a neural network to act as a powerful speaker embedding extractor since one must address capturing speaker information from both local and global patterns in speech signals. There also several requirements for a networks to be applicable in real tasks: its computational lightness and high generalization. Therefore we consider two alternative architectures.

\subsection{Max pooling embeddings}
\label{sec:max-poolemb}

We introduce a modification of the original TDNN-based x-vector extractor \cite{snyder2017deep} with intermediate max pooling layers, which we call a max pooling embedding extractor. Max pooling layers are very common within convolutional neural networks and are a source of spatial invariance and dimensionality reduction of data for algorithms in addition to necessary non-linearity. Including intermediate max pooling layers effectively turns a TDNN architecture to a more CNN-like.
It reduces amount of network parameters and speeds up computations.

Time-delay layers use PReLU \cite{he2015delving} instead of the conventional ReLU activation function.
For the segment level of the network, which comes after the stats pooling layer, we also make use of an alternative activation function. Here we use Max-Feature-Map (MFM) activation \cite{lavrentyeva2017audio} in place of ReLU. In contrast to commonly used ReLU function that suppresses neurons by a zero threshold, MFM suppresses neurons by mutually competitive relationships. By doing so the MFM activation acts as an embedded feature selector. 



Our x-vector modification architecture is decribed in Table~\ref{tab:maxdnn}. Max pooling is made with windows are $2 \times 2$ with stride $2$. $N_\text{spk}$ is the number of speakers in training set. Layer context can be thought as of width of a convolution filter along time dimension if expressed in terms of CNNs.

\begin{table}[ht]
    \caption{Max pooling embedding extractor configuration. Frame layers correspond to the TDNN architecture part of the network, while segment layers to the fully-connected one. Stats pooling layer is the intermediate time-folding layer.}
    \label{tab:maxdnn}
    \centering
    \begin{tabular}{l|c|c|c}
        \hline
         \textbf{Layer} & \textbf{\thead{Layer \\ context}} & \textbf{\thead{Total \\ context}} & \textbf{In $\times$ out}\\
        \hline
        frame1 & 7 & 7 & 161 $\times$ 256 \\
        maxpool1 & 2 & 8 & 256 $\times$ 128\\
        frame2 & 5 & 18 & 640 $\times$ 256 \\
        maxpool2 & 2 & 20 & 256 $\times$ 128\\
        frame3 & 3 & 28 & 384 $\times$ 256 \\
        maxpool3 & 2 & 32 & 256 $\times$ 128\\
        frame4 & 1 & 32 & 256 $\times$ 2048 \\
        maxpool4 & 2 & 32 & 2048 $\times$ 1024\\
        stats pooling & all & all & 1024 $\times$ 2048 \\
        segment6 MFM & all & all & 2048 $\times$ 1024 \\
        segment7 MFM & all & all & 1024 $\times$ 512 \\
        A-softmax  & all & all & 512 $\times$ $N_\text{spk}$ \\
        \hline
    \end{tabular}
\end{table}

It can be seen that frame-level layers are responsible for capturing time-local speaker features, while stats pooling collects global information.

\subsection{Deep residual embeddings} \label{sec:resemb}

The necessity of wide context capturing seems essential when trying to collect informative speaker features and separate them from other signal variations.
There are two ways of context expanding in TDNN architecture:
either by widening it at each frame-level layer,
or by deepening the network to accumulate richer context with higher level of feature abstraction. 

\begin{figure*}[h]
\centering
\includegraphics[width=0.7\textwidth]{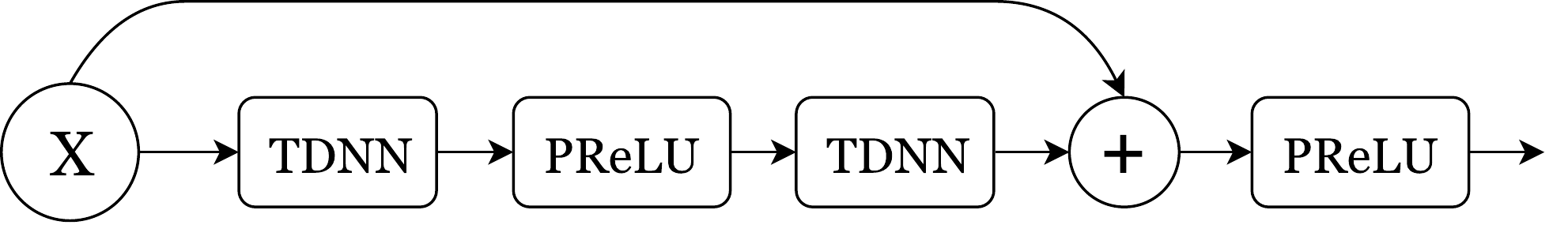}
\caption{TDNN Residual block}
\label{fig:resblock}
\end{figure*}

Our second alternative architecture is a deep extractor consisting of time-delay layers with shallow frame-level contexts. A practical way of training a rather deep network is additional of residual connections, which were initially introduced in paper~\cite{he2016deep} for deep image representation learning.
Hence we introduce a residual TDNN block, illustrated by Figure~\ref{fig:resblock}.
A technical detail is that we need to ensure that the dimensions match when adding residual part to output in a residual block. For this purpose zero padding was applied.

A big advantage of using residual connections is the ability to pass primitive features to top layers. The network itself adapts to the level of abstraction needed at each layer by leveraging weights, and so the effective width of the context for each level is also adapted from the speech setting.

Our deep residual TDNN extractor contains $M$ residual blocks, a stats pooling layer for feature aggregating through time and a fully-connected classifier.
The network architecture is described precisely in Table~\ref{tab:resnet}.

\begin{table}[t]
    \centering
    \caption{Deep residual embedding extractor configuration. $N_{spk}$ is the number of speaker classes, which determines the number of neurons at the output layer.}
    \label{tab:resnet}
    \begin{tabular}{l|c|c|c}
        \hline
         \textbf{Layer} & \textbf{\thead{Layer \\ context}} & \textbf{\thead{Total \\ context}} & \textbf{In $\times$ out}\\
        \hline
        frame1 & 3 & 3 & 69 $\times$ 128 \\
        maxpool1 & 2 & 4 & 128 $\times$ 64\\
        \hline
        \thead[l]{frame2: \\ resTDNN block 1} & 3 & 8 & 64 $\times$ 64 \\
        \hline
        \textellipsis \\
        \hline
        \thead[l]{frame $M+1$: \\ resTDNN block $M$} & 3 & $8M$ & 64 $\times$ 64 \\
        \hline
        frame $M+2$ & 1 & $8M$ & 64 $\times$ 2048 \\
        maxpool $M+2$ & 2 & $16M$ & 2048 $\times$ 1024\\
        \hline
        stats pooling & all & all & 1024 $\times$ 2048 \\
        segment6 MFM & all & all & 2048 $\times$ 1024 \\
        segment7 MFM & all & all & 1024 $\times$ 512 \\
        A-softmax  & all & all & 512 $\times$ $N_\text{spk}$ \\
        \hline
    \end{tabular}
\end{table}

\subsection{Backends} \label{sec:backends_desc}

In our experiments we measured quality of speaker embeddings with a backend and without one.
In the last case simple cosine similarity metric can be applied for verification directly in the embedding space.
Moreover, the discriminative metric learning approach could be viewed as a scalable alternative to simple cosine metric or LDA-PLDA backend for deep speaker embeddings.



In this work we study how Cosine Similarity Metric Learning can improve embedding verification only for x-vectors based system. In CSML, a linear transformation $\matr{A}$ must be learned to compute cosine similarities (CS) on a pair $(\vect{x_1}, \vect{x_2})$ as follows:   

\begin{equation}
\mathcal{S(\vect{x_1},\vect{x_2}, \matr{A})} = \dfrac{(\matr{A}\vect{x_1})^T(\matr{A}\vect{x_2})}{{\norm{\matr{A}\vect{x_1}}}{\norm{\matr{A}\vect{x_2}}}}
\end{equation}
where the transformation matrix $\matr{A} $ is upper triangular. Under this constraint $\matr{A}^T\matr{A}$ is positive-definite.  
Unlike \cite{Nguen} we set the triplet loss objective function for training $\matr{A}$:

\begin{equation}
\mathcal{L(\matr{A})}=  \sum_{a,p,n\in{T}}\log(1+\exp(-d_{a,p,n}))\\
\end{equation}
where $ d_{a,n,p} = s_{a,p} - s_{a,n}$ is the difference between similarity scores $ s_{a,p}$  and  $s_{a,n}$. 
$T$ is a collection of training triplets which is formed from a training dataset. A triplet $(a, p,  n)$ contains an anchor sample  $a$ as well as a positive $p \neq a $  and a negative $n$ example of the anchor's identity. We included in $T$ all positive examples and only 1500 hardest negative examples for selected anchors.

\section{Implementation details}
\label{sec:frontend_backend}

This section describes details of the studied speaker recognition systems.


\subsection{DNN posterior-based i-vectors}

We extracted DNN posterior-based i-vectors as described in \ref{sec:dnn_posterior} for a baseline. A DNN was trained on the Switchboard corpus using the Kaldi speech recognition toolkit \cite{povey2011kaldi}. Outputs of the DNN correspond to the set of 2700 speech triphone states. These 2700 speech-related outputs were used for statistics calculation.
20 MFCCs (including log energy) were calculated using 23 filter banks with their first and second derivatives. Mean and variance normalization was subsequently applied.

This system is called DNN/i-vector.

\subsection{X-vectors}
\label{sec:xvec_desc}

We used the same x-vector based system configuration as described in \cite{snyder2017deep} and available in Kaldi \cite{povey2011kaldi}. 
Our system takes 23 dimensional MFCCs as input instead of raw filterbanks. MFCCs are computed  with a frame-length of 25ms, mean-normalized over a 3-second sliding window. Energy-based voice activity detector (VAD) filters out non-speech frames.

The speaker embeddings are extracted from the second to last layer of the classifier network. We applied Cosine Similarity Metric Learning approach (see \ref{sec:backends_desc}) instead of a LDA-PLDA backend as an alternative embedding classifier.

We refer to this system as X-vectorNet.

\subsection{Speaker Max-Pooling Net}
\label{sec:maxpoolnet}

Our first development is a max pooling TDNN-based speaker embedding extractor, which is described in Section \ref{sec:max-poolemb}.
Max-Pooling TDNN uses the same front-end features as x-vector based system in section~\ref{sec:xvec_desc}. 
The network is trained on short segments of speech (3-10 sec), which are randomly sampled from the training data.

After classificator training, the last fully-connected layer with its angular softmax activation is removed from the network in order to obtain an extractor of high-level representations for speaker specific information. Simple cosine metric as well as LDA-PLDA approach can be used as a backend to the Max-pooling TDNN embeddings.

The model was implemented in PyTorch \cite{paszke2017automatic} and trained using a single GeForce GTX 1080 GPU.

We refer to this system as SpeakerMaxPoolNet.

\subsection{Speaker Residual Net}

Our deepest embedding extractor architecture is represented by a deep neural network with TDNN residual blocks, as described in Section \ref{sec:resemb}. It takes the same MFCC input features as X-vectorNet and SpeakerMaxPoolNet.
The network is trained on short random segments of speech (3-10 sec), which are randomly sampled from the training data.

The embeddings are also extracted from the penultimate fully-connected layer, as it is done in \ref{sec:maxpoolnet}.
Again, cosine metric as well as LDA-PLDA approach can be used as a backend to the Speaker Residual Net speaker embeddings.

The model was also implemented in PyTorch \cite{paszke2017automatic} and trained using a single GeForce GTX 1080 GPU.

We refer to this system as SpeakerResNet.

\section{Experimental setup}
\label{sec:setup}

\subsection{Training data}

We have prepared multiple training settings during our series of experiments.
For preliminary studies, we used NIST 1998-2008 datasets for training with no data augmentation.

In our main experimental setup, telephone speech is collected as training data. It includes Switchboard2 Phases 1, 2, and 3, Switchboard Cellular and data from NIST SREs from 2004 through 2010. In addition, we use data augmentation as it was done in \cite{snyderx} to increase amount and diversity of the training data. In total, there are about 55,000 recordings from 5,277 speakers in this training part, a major part of which is English speech. We refer to this data as \textit{English} data.

In other experiments we also used Russian speech subcorpus named RusTelecom to extend the training set. RusTelecom is a proprietary Russian speech corpus of telephone speech, collected by call-centers in Russia.
The train part of the RusTelecom database consists of approximately 70000 sessions from 11087 speakers. We refer to this data as \textit{Russian} data.

\subsection{Evaluation}

For preliminary studies, we used NIST 2010 evaluation dataset for testing under the det5 protocol.

Our main experimental setup includes evaluation on the Speaker-in-the-Wild \cite{mclaren20162016} (SITW) and NIST SRE 2016 \cite{nist2016eval} datasets. In the case of NIST SRE 2016 we used the unequalized protocol.

We report results in terms of equal error rate (EER) and the minimum detection cost function (DCF) with $P_\text{Target}=10^{-2}$ and $P_\text{Target}=10^{-3}$. 

\section{Experimental results}
\label{sec:results}


\subsection{Preliminary investigation}
\subsubsection{Angular Softmax vs regular Softmax}

\begin{figure}[h]
  \centering
  \includegraphics[width=0.9\linewidth]{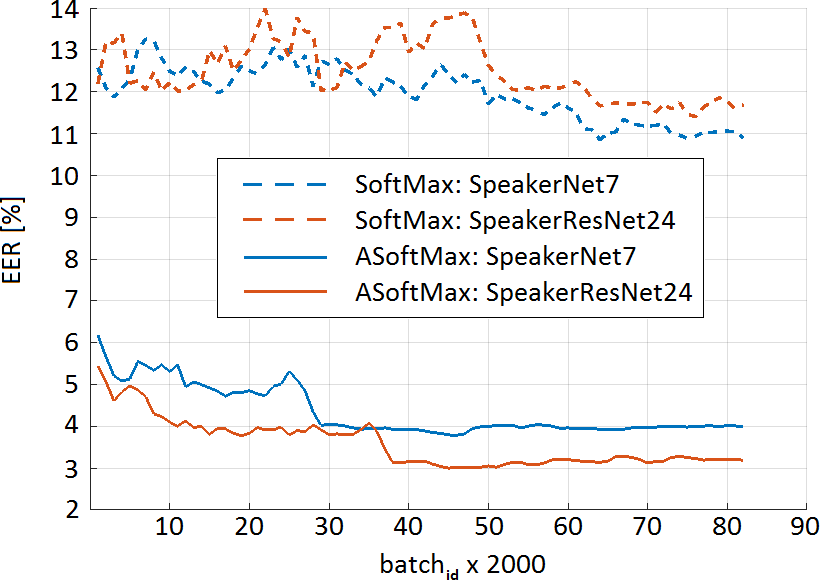}
  \caption{Comparison of speaker recognition performance on NIST 2010 det5 protocol for softmax and a-softmax classifiers used during traning. The numbers in the labels indicate the total amount of layers in the extractor.}
  \label{fig:sxvsasx}
\end{figure}

We compare effectiveness of the regular softmax and the margin-based A-softmax cross-entropy losses in training a deep speaker embedding extractor.
Figure~\ref{fig:sxvsasx} shows NIST 2010 det5 protocol EER to the number of training iterations. We experiment with the two proposed architecture solutions, namely SpeakerMaxPoolNet and SpeakerResNet.

These results demonstrate lack of generalization when using regular softmax for training. In contrast, margin-based A-softmax objective leads to comparatively good speaker generalization in the obtained discriminative speaker embedding space. It should be noted that simple cosine scoring was used for calculating system performance. Application of more complex backends such as LDA-PLDA slightly improves the results for embeddings trained with the regular softmax.

\subsubsection{"Clean" condition experiments}

The preliminary experiment results in the "clean" telephone speech conditions are presented in Table~\ref{tab:nists}. The NIST2010 evaluation protocol was used for testing. Again, we focus on SpeakerMaxPoolNet and SpeakerResNet architectures. A-softmax margin cross-entropy loss was applied to train the networks.

\begin{figure}[h]
  \centering
  \includegraphics[width=0.9\linewidth]{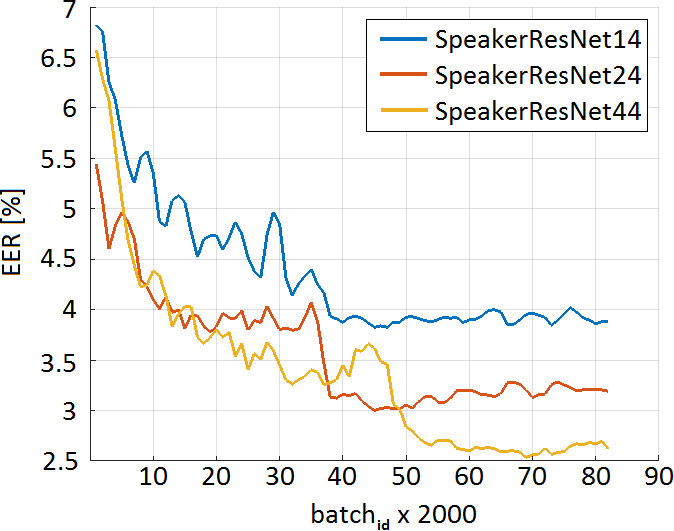}
  \caption{EER evolution on NIST 2010 det5 protocol for different architectures. The numbers in the labels indicate the total amount of layers in the extractor.}
  \label{fig:eerresnet}
\end{figure}

\begin{figure}[h]
  \centering
  \includegraphics[width=0.9\linewidth]{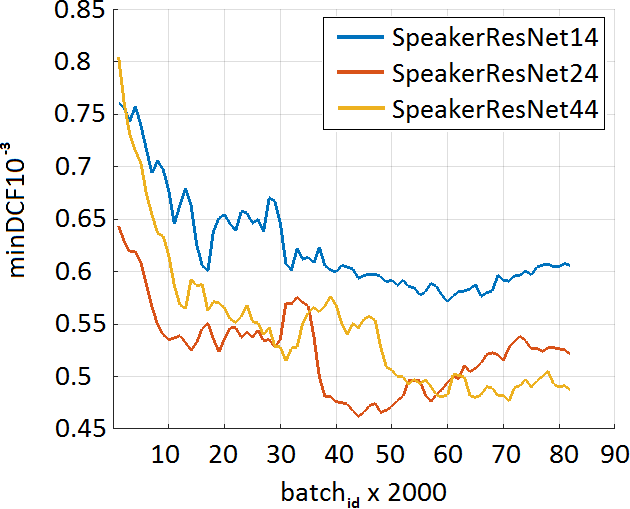}
  \caption{MinDCF$10^{-3}$  evolution on NIST 2010 det5 protocol for different architectures. The numbers in the labels indicate the total amount of layers in the extractor.}
  \label{fig:mindcfresnet}
\end{figure}

Figures~\ref{fig:eerresnet},~\ref{fig:mindcfresnet} illustrate EER and minDCF$^{-10}$ evolution during DNN training. The results in Table~\ref{tab:nists} and Figures~\ref{fig:eerresnet},~\ref{fig:mindcfresnet} show that 
deep networks with residual frame-level connections are superior to wide but relatively shallow architectures. Unfortunately, these systems were not able to surpass DNN/i-vector baseline system in terms of quality in "clean" conditions.  

\begin{table}[]
    \centering
    \caption{NIST 2010 det5 protocol evaluation results.}
    \label{tab:nists}
    \begin{tabular}{c|c|c|c}
        \hline
        \textbf{System}& \textbf{Backend} & EER, \% & DCF$10^{-3}$ \\
        \hline
         DNN/i-vector & \thead{LDA-\\PLDA} & 1.67 & 0.3415 \\
         SpeakerMaxPoolNet7 & cos & 3.65 &  0.5866 \\
         SpeakerMaxPoolNet7 & \thead{LDA-\\PLDA} & 3.71 & 0.5836 \\
         SpeakerResNet24 & cos & 3.01 & 0.498 \\
         SpeakerResNet24 & \thead{LDA-\\PLDA} & 3.14 & 0.5131 \\
         SpeakerResNet44 & cos & 2.72 & 0.4967 \\
         SpeakerResNet44 & \thead{LDA-\\PLDA} & 2.76 & 0.5256 \\
         \hline
    \end{tabular}
\end{table}



\begin{table*}[]
    \centering
    \begin{tabular}{c|c|c|c|c|c|c|c}
        \hline
        \multirow{2}{*}{\textbf{System}} &
        \multirow{2}{*}{\textbf{Backend}} &
        \multicolumn{3}{c|}{NIST2016} & \multicolumn{3}{c}{SITW} \\
        \cline{3-8}
         && EER, \% & DCF$10^{-2}$ & DCF$10^{-3}$ & EER, \% & DCF$10^{-2}$ & DCF$10^{-3}$ \\
        \hline
         \thead{X-vectorNet} & LDA-PLDA & 16.88 &  1.00 & 1.00
         & 7.82 & 0.6136 &  0.7753\\
         \thead{X-vectorNet} & CSML & \textbf{13.26} & \textbf{0.8686} & 0.9972 & 8.58 & 0.5916 & 0.7487\\
         \thead{SpeakerMaxPoolNet7 }& cos & 14.57 & 0.9340 & 0.9930& 7.72 & 0.5573 & 0.7320 \\
         \thead{SpeakerResNet24} & cos & 14.18 & 0.9257 & \textbf{0.9881} & \textbf{6.72} & \textbf{0.5200} & \textbf{0.7156} \\
         \hline
    \end{tabular}
    \caption{Results using \textit{English} corpora for training. No adaptation implemented.}
    \label{tab:eng}
\end{table*}

\begin{table*}[]
    \centering
    \begin{tabular}{c|c|c|c|c|c|c|c}
        \hline
        \multirow{2}{*}{\textbf{System}} &
        \multirow{2}{*}{\textbf{Backend}} &
        \multicolumn{3}{c|}{NIST2016} & \multicolumn{3}{c}{SITW} \\
        \cline{3-8}
         && EER, \% & DCF$10^{-2}$ & DCF$10^{-3}$ & EER, \% & DCF$10^{-2}$ & DCF$10^{-3}$ \\
        \hline
         \thead{X-vectorNet} & LDA-PLDA & 11.89 &  0.85 & 1.00
         & 6.57 & 0.6031 & 0.7870\\
         \thead{X-vectorNet} & CSML & \textbf{10.97} & \textbf{0.7025} & 0.9294 & 8.06 & 0.5947 & 0.7490\\
         \thead{SpeakerMaxPoolNet7 }& cos & 11.50 & 0.7545 & 0.9241& 6.78 & 0.5413 & 0.7152 \\
         \thead{SpeakerResNet24} & cos & 11.13 & 0.7332 & \textbf{0.8963} & \textbf{6.18} & \textbf{0.5079} & \textbf{0.7066} \\
         \hline
    \end{tabular}
    \caption{Results using \textit{English} corpora for training. Centering on in-domain devset implemented.}
    \label{tab:eng_adapted}
\end{table*}

\begin{table*}[]
    \centering
    \begin{tabular}{c|c|c|c|c|c|c|c}
        \hline
        \multirow{2}{*}{\textbf{System}} &
        \multirow{2}{*}{\textbf{Backend}} &
        \multicolumn{3}{c|}{NIST2016} & \multicolumn{3}{c}{SITW} \\
        \cline{3-8}
         && EER, \% & DCF$10^{-2}$ & DCF$10^{-3}$ & EER, \% & DCF$10^{-2}$ & DCF$10^{-3}$ \\
        \hline
         \thead{X-vectorNet} & LDA-PLDA & 15.03 &  0.9974 & 1.00
         & 11.62 & 0.7722 & 0.8971 \\
         \thead{X-vectorNet} & CSML & \textbf{12.87} & \textbf{0.8602} &  0.9894 & 9.62 & 0.6318 & 0.7861\\
         \thead{SpeakerMaxPoolNet7 }& cos & 13.09 & 0.8811 & 0.9879 & 7.35 & 0.5768 & 0.7585 \\
         \thead{SpeakerResNet24} & cos & 13.94 & 0.8937 & \textbf{0.9869} & \textbf{7.08} &  \textbf{0.5351} & \textbf{0.7025} \\
         \hline
    \end{tabular}
    \caption{Results using \textit{English} and \textit{Russian} datasets for training. No adaptation implemented.}
    \label{tab:engrus}
\end{table*}

\begin{table*}[]
    \centering
    \begin{tabular}{c|c|c|c|c|c|c|c}
        \hline
        \multirow{2}{*}{\textbf{System}} &
        \multirow{2}{*}{\textbf{Backend}} &
        \multicolumn{3}{c|}{NIST2016} & \multicolumn{3}{c}{SITW} \\
        \cline{3-8}
         && EER, \% & DCF$10^{-2}$ & DCF$10^{-3}$ & EER, \% & DCF$10^{-2}$ & DCF$10^{-3}$ \\
        \hline
         \thead{X-vectorNet} & LDA-PLDA & 12.30 &  0.8732 & 1.00
         & 11.73 & 0.7802 &  0.8984\\
         \thead{X-vectorNet} & CSML & \textbf{10.45} & \textbf{0.6914} & 0.9235 & 8.70 & 0.6216 & 0.7998\\
         \thead{SpeakerMaxPoolNet7 }& cos & 11.26 & 0.7311 & 0.9241 & 6.40 & 0.5397 & 0.7236\\
         \thead{SpeakerResNet24} & cos & 11.16 & 0.7128 & \textbf{0.9024} & \textbf{5.90} & \textbf{0.5125} & \textbf{0.6987}\\
         \hline
    \end{tabular}
    \caption{Results using \textit{English} and \textit{Russian} datasets for training. Centering on in-domain devset implemented.}
    \label{tab:engrus_adapted}
\end{table*}

\subsection{Main experiments} \label{sec:main_exp}
In the main experiments we tested speaker recognition systems against "in the wild" conditions.
To do this we trained DNN extractors on large augmented corpora. We tested two alternative extractors: SpeakerMaxPoolNet with 7 layers and SpeakerResNet with 24 layers.

According to the results of \cite{snyder2017deep,snyderx}, x-vector based systems with LDA-PLDA model backend are superior to standard i-vector based systems. Thus we used x-vector LDA-PLDA solution as a baseline in this case. Note that we did not used any embedding adaptation methods apart from centering on in-domain development set.

Tables~\ref{tab:eng},~\ref{tab:eng_adapted} present the results for systems trained only on \textit{English} corpora. One can observe that SpeakerMaxPoolNet7 and SpeakerResNet24 perform well, despite simple cosine similarity scoring was applied. Moreover, the systems which were trained with margin angular softmax loss function outperform the x-vector baseline.
We observe that the x-vector based LDA-PLDA system needs in-domain centering more than the other studied systems (see Table~\ref{tab:eng_adapted}). Also, using Cosine Similarity Metric Learning backend leads to significant performance improvement in comparison to LDA-PLDA.  

Tables~\ref{tab:engrus},~\ref{tab:engrus_adapted} show system performance when an extended dataset containing \textit{English} and \textit{Russian} corpora is used for training. A notable average performance improvement for the SpeakerMaxPoolNet7 and SpeakerResNet24 based systems on both SITW and NIST 2016 evaluation is seen. In contrast, the x-vector based systems experience some degradation on SITW protocol.





%
%

\section{Discussion}
\label{sec:discuss}
In our investigations we explored different strategies for discriminative speaker extractor training on the closed set task.
We found that first-order optimization for regular softmax objective such as stochastic gradient descend always leads to fast overfitting on the closed set speakers, and
extractors obtained this way generalize badly for open-set task.

According to our observations, one reason of the x-vector success is the natural-gradient (NG) modification \cite{povey2014parallel} of the stochastic gradient descend (SGD) optimization. The natural-gradient SGD procedure uses an inverse Fisher matrix for gradient scaling during training and prevents convergence to local optima.

We found out that the choice of the optimization objective is essential for training a discriminative speaker recognition system.
According to our preliminary experiment results, angular margin softmax loss is much more effective than regular softmax loss as a discriminative objective. We did not try to apply A-softmax loss for x-vector extractor training but we believe that it can improve discriminative properties of the x-vectors.

Our studies also allow us to conclude that the performance of x-vector based systems can be improved by using cosine similarity metric learning method as a backend.

Another major issue of the speaker verification problem is the choice of the DNN architecture. The results presented in Tables~\ref{tab:eng}, \ref{tab:eng_adapted},~\ref{tab:engrus},~\ref{tab:engrus_adapted} demonstrate good performance of the proposed alternative extractors. Deep context based architecture SpeakerResNet with 24 layers is superior to SpeakerMaxPoolNet in real life conditions.
We speculate that residual TDNN connections allow the network to automatically leverage necessary context for each level of feature abstraction.

Simple cosine similarity scoring method can be used for speaker verification in these systems.
We've decided not to use CSML with SpeakerResNet and MaxPoolNet since A-softmax loss is specifically designed for use with cosine similarity and trains the last network layers accordingly.

When trained with augmented data, DNN-based speaker embedding systems significantly outperform our previous i-vector-based systems on SITW protocol \cite{kudashev2016speaker}.



\section{Conclusion}
\label{sec:conclusion}
This work demonstrates that DNN-based speaker embedding extractors can be effectively used for speaker verification.

\begin{itemize}
    \item Choice of the optimization objective is essential for obtaining speaker embedding extractor with good generalization properties. To effectively discriminate speaker embeddings by cosine similarity a-softmax can be used.
    \item Speaker features are best captured with deep context. Using rather deep TDNN architectures with residual connections can capture speaker features from any level of feature abstraction and context.
    \item Performance of x-vector based systems could be improved by using cosine similarity metric learning approach for a backend model.
\end{itemize}

We proposed two speaker embedding extractors called SpeakerMaxPoolNet and SpeakerResNet which performed well during evaluation on SITW and NIST SRE 2016 corpora.

\section{Acknowledgements}

This work was financially supported
by the Ministry of Education and Science of the Russian
Federation, Contract 14.578.21.0189
(ID RFMEFI57816X0189).

\bibliographystyle{IEEEbib}

\end{document}